\documentstyle[12pt,epsfig]{article}
\def\be{\begin{equation}} \def\ee{\end{equation}} \def\bea{\begin{eqnarray}}
\def\eea{\end{eqnarray}} \def\nnb{\nonumber}
\def\calO{{\cal{O}}}

\begin{document}

\hfill{TRI-PP-04-10}

\begin{center}
\vskip 4mm
{\Large\bf
Effective field theory of the deuteron \\
with dibaryon field
}
\vskip 1cm
{\large Shung-ichi Ando $^{a,}$\footnote{E-mail:sando@triumf.ca}
and Chang Ho Hyun $^{b,\, c,}$\footnote{E-mail:hch@meson.skku.ac.kr}
}\\
\vskip 7mm
{\it $^a$ Theory Group, TRIUMF, 4004 Wesbrook Mall, Vancouver,
B.C. V6T 2A3, Canada}\\
\vskip 2mm
{\it $^b$ School of Physics, Seoul National University,
Seoul 151-742, Korea}\\
\vskip 2mm
{\it $^c$ Institute of Basic Science, Sungkyunkwan University,
Suwon 440-746, Korea}
\end{center}

\vskip 1.5cm
Pionless effective field theory
with dibaryon fields
is reexamined for observables involving the deuteron.
The electromagnetic form factors of the deuteron
and the total cross section of
radiative neutron capture on the proton,
$n\, p \rightarrow d\, \gamma$, are calculated.
The low energy constants
of vector(photon)-dibaryon-dibaryon vertices
in the effective lagrangian
are fixed primarily
by the one-body vector(photon)-nucleon-nucleon
interactions. 
This scheme for fixing the values of the low energy constants
satisfactorily reproduces the results of
the effective range theory.
We also show that,
by including higher order corrections,
one can obtain results
that are close to those of 
Argonne v18 potential model.


\newpage

\noindent
{\bf 1. Introduction}

Effective field theory (EFT)
has proven to provide a useful tool for describing
a wide class of meson-meson, meson-nucleon,
and nucleon-nucleon ($NN$) processes
with and without external probes
in the low energy regime~\cite{w-97}.
However, special care for the $NN$ processes
is required to deal with
the $^1 S_0$ channel for its long scattering length
(or existence of a quasi bound state) and 
the $^3 S_1 - {}^3 D_1$ channel 
for existence of the deuteron. 
In general, EFT is based on
the perturbative expansion of
physical observables in terms of
small external momenta,
but a non-perturbative treatment 
would be required for
positions of the singularity at
the small energy scales, 
which associated with the long scattering length 
and the small binding energy, 
compared to the chiral symmetry breaking scale, 
pion mass $m_\pi$.\footnote{
Even if the scattering lengths 
were of the order of pion mass, still 
nonperturbative treatment 
of the $NN$ amplitude would be required. 
See Ref.~\cite{bbsk-npa02}.}
To deal with the problem,
Weinberg has suggested counting rules
that allow one to handle this 
non-perturbative problem
and derive the $NN$ potential systematically~\cite{w-plb90}.
With the potential calculated to a given order,
the $S$-matrix is calculated from a wave
function obtained by solving
the Schr\"{o}dinger equation. 
This scheme has shown good accuracy and
convergence with only
a few leading terms~\cite{ork-prc96,pkmr-npa99,hpm-plb00,egm-npa00}.
Kaplan, Savage, and Wise (KSW) suggested
new counting rules, in which pions are treated perturbatively
and do not appear in leading order (LO). 
They also employed the power divergence
subtraction scheme for 
regularization~\cite{ksw-plb98}. 
Only a LO contact two-nucleon interaction
is treated non-perturbatively
and thus physical observables can be directly 
calculated from 
Feynman diagrams expanded order by order.
Over the last decade both the Weinberg and 
KSW schemes have been used extensively
in studying few-nucleon systems.\footnote{ 
Detailed and careful studies of both of the schemes 
have been reported in Refs.~\cite{bbsk-npa02,fms-npa00}.}
For reviews, see, 
{\it e.g.}, Refs.~\cite{bbhps-00} and \cite{bk-02}.

Convergence of deuteron observables
becomes slow near the deuteron pole
due to a large expansion parameter
$\gamma\, \rho_d\simeq 0.4$,
where $\gamma \simeq 45.7$ MeV and $\rho_d \simeq 1.764$ fm
in the KSW scheme.
This would be a typical expansion parameter of
the deuteron channel both in
pionless effective field theory
(pionless EFT)~\cite{vk-npa99,g-98},
where the pions are regarded as
heavy degrees of freedom in study of very low energy
reactions and integrated out from effective lagrangian,
and in theories that treat pions perturbatively.
It was suggested, however,
that adjusting the deuteron wave function
to fit the asymptotic $S$-state normalization constant,
$Z_d = \gamma\, \rho_d/(1 - \gamma\, \rho_d)$,
assures more efficient convergence~\cite{rho99,pcnpa00,prs-plb00}.
By introducing a dibaryon field
which represents a resonance state (or a bound state)
of two nucleons\footnote{
Initially,
the dibaryon field was introduced by Kaplan
in the $^1S_0$ channel $NN$ scattering~\cite{k-npb97}.
Subsequently, 
the dibaryon formalism without pions
was applied to studies of three nucleon systems 
by Bedaque and van Kolck\cite{bk-plb98}. 
It was shown that the pole structure of the amplitude
at the low energies can be more efficiently
reproduced with the use of the dibaryon field
and  resummation of effective range.
},
Beane and Savage showed that
in terms of dibaryon EFT (dEFT) without pions,
a long tail of the deuteron wave function
can be naturally derived from
the two-nucleon(dibaryon) propagator
at the deuteron pole~\cite{bs-npa01}.
It was also shown that, with the use of dEFT,
a good number of diagrams at a given order
even in pionless EFT 
can be cast into a few diagrams. 
This feature is expected to be useful
when higher order corrections need to be calculated.
One can also expect that dEFT should be
an attractive approach
in attempts to incorporate,
in the framework of EFT,
radiative corrections
(photon degrees of freedom)
for the two-nucleon reactions
~\cite{bp-prd01,afgkmsn-plb04}.

In applying dEFT to reactions 
involving an external probe,
there appear additional 
low energy constants (LEC's)
in effective lagrangian.
LEC's that enter into electromagnetic (EM)
interaction on the two-nucleon system 
appear in the vector(photon)-dibaryon-dibaryon 
($Vdd$) interactions 
in terms of the dibaryon fields
and can be fixed, e.g., from
the total cross section of radiative
neutron capture on a proton,
$np \rightarrow d \gamma$,
or from the deuteron EM multipole moments.\footnote{
Recently Detmold and Savage \cite{ds-04}
suggested that lattice simulations allow us
to estimate the LEC's that feature
in dEFT describing reactions with an external probe.}
These LEC's are considered to subsume
the high energy physics,
such as meson exchange currents or
heavy meson exchanges,
that has been integrated out from the effective lagrangian.
Using Weinberg's counting rule with
pionful theory, these contributions 
from the meson exchange currents
are higher order corrections 
and are known that their corrections
are numerically about a few $\sim$ 10\% 
in the reactions considered in this work.
Similar results have been reported
in the KSW scheme and pionless EFT
(the references will be given in each section.)
However, as we will show, 
the contributions from the LEC's of 
the $Vdd$ vertices in dEFT
turn out to be anomalously larger
(about 40 $\sim$ 50\% of LO contributions)
than the former EFT results.

Each version of EFT has its own expansion scheme
and values of LEC's for a given vertex 
can differ from one version to another.
The large ($40 \sim 50\%$) cotributions 
from the LEC terms in dEFT 
may be regarded as consistent 
with the expansion parameter 
($\gamma\rho_d \simeq 0.4$)
in the pionless theories \cite{vk-npa99,g-98}.
However, such large contributions 
from the LEC terms with the EM probe 
could alter our knowledge about this class of the LEC's 
and imply that convergence can be much slower than 
that found in the previous EFT calculations. 
This could be a serious drawback
in calculations with dEFT,
since to obtain sufficient accuracy, 
one requires a higher order calculation
due to the slow convergence.
Subsequently, 
taking into account one higher order 
can be a formidable task
and more importantly,
the slow convergence of dEFT 
can make the uncertainty of the results 
at a given order larger than 
that in the other EFT calculations.
In order to make the convergence of dEFT more efficient,
we propose a simple way to handle this class of the LEC's,
{\it i.e.} the LEC's of the $Vdd$ vertices, in dEFT.

In this work, we explore several applications
of dEFT for the deuteron reactions.
We consider a pionless effective lagrangian
that includes the dibaryon fields in the
${}^1S_0$ and ${}^3S_1-{}^3D_1$ channels.
Those LEC's appearing in the strong interaction part of the
dibaryon lagrangian are fixed
by the effective range parameters in each channel.
The EM interactions between the dibaryon fields and
an external vector field
introduce additional LEC's 
in both isoscalar and isovector transitions.
At first, we fix
the LEC's of the $Vdd$ vertices 
by experimental data.
The contributions from them, 
as mentioned before,
turn out to be anomalously larger than those
found in the previous EFT calculations.
We suggest modified counting rules in which these LEC's are
enhanced by an order of $Q^{-1}$.
In the new counting scheme, we suppose that
a prime portion of the LEC's of the $Vdd$ vertices in dEFT
originates mainly from 
one-body vector(photon)-nucleon-nucleon ($VNN$) interactions.
We find that this scheme for fixing the LEC's 
makes the convergence of dEFT as efficient as other EFT
and well reproduces the deuteron EM form factors 
and the total cross section for
$np\to d\gamma$ 
obtained in the effective range theory (ERT)~\cite{b-pr49}.
Including higher order terms in our dEFT calculations,
we find that the results of dEFT become
comparable to those of 
potential model calculation using
Argonne v18 potential\cite{wss-prc95}.

This paper is organized as follows:
In Sect.~2 the effective lagrangian is given.
We calculate in Sect.~3
the EM form factors of the deuteron
and fix the LEC's in the EM interaction lagrangians.
We then calculate physical observables 
in the elastic $e$-$d$ scattering, and
compare our results with those of ERT
and a potential model calculation, 
and also with the experimental data. 
In Sect.~4, we investigate the $np\to d\gamma$ reaction.
The final section, Sect.~5, 
is devoted to discussion and conclusions.
In Appendix, we give
details on the two-nucleon(dibaryon) propagators
and $NN$ scattering amplitudes,
and determine the LEC's in the strong-interaction lagrangian.

\vskip 3mm \noindent
{\bf 2. Effective lagrangian with dibaryon fields}

A pionless effective lagrangian 
for the nucleon and the dibaryon fields 
interacting with an external vector field 
can be written as
\bea
{\cal L} = {\cal L}_N
+ {\cal L}_s
+ {\cal L}_t
+ {\cal L}_{st},
\eea
where ${\cal L}_N$ is the one-nucleon lagrangian,
${\cal L}_s$ and ${\cal L}_t$ are the lagrangians for the
dibaryon fields in the ${}^1S_0$ and ${}^3S_1$ channels,
respectively.
${\cal L}_{st}$ is
the lagrangian that accounts for the transition between
${}^1S_0$ and 
${}^3S_1$ channels
through the isovector EM interaction.\footnote{
The counting rule we employ here 
is the same as that in Ref.~\cite{bs-npa01}:
we do not expand the amplitudes 
in terms of the effective range,
$\rho_d$ or $r_0$.
Introducing an expansion scale 
$Q < \Lambda (\simeq m_\pi)$, 
we count magnitude of spatial part of the external 
and loop 
momenta,
$|\vec{p}|$ and $|\vec{l}|$, as $Q$,
and the time component of them, $p^0$ and $l^0$, 
as $Q^2$. 
Thus the nucleon and dibaryon propagators
are of $Q^{-2}$ and a loop integral leads to $Q^5$.
The scattering lengths and effective ranges are
counted as $Q\sim \{\gamma,1/a_0, 1/\rho_d,1/r_0\}$.
Orders of vertices and diagrams are easily 
obtained by counting the numbers of these factors.
LEC's are counted as $Q^0$ initially. However we will argue 
that some of them 
(the $Vdd$ vertices)
can be enhanced by $Q^{-1}$.
\label{footnote;counting}
} 

${\cal L}_N$ in the heavy-baryon formalism reads
\bea
{\cal L}_N &=&
N^\dagger \left\{
iv\cdot D
+\frac{1}{2m_N}\left[
(v\cdot D)^2-D^2 -i[S^\mu,S^\nu]
\left((1+\kappa_V)f^+_{\mu\nu}
+(1+\kappa_S)f^S_{\mu\nu}\right)
\right]
\right. \nnb \\ && \left.
+ 2 \tilde{d}_7 v^\nu [D^\mu,f_{\mu\nu}^S]
+ 2 \tilde{e}_{54} i[S^\mu,S^\nu] [D^\lambda,[D_\lambda, f_{\mu\nu}^S]]
+ \cdots 
\right\} N\, ,
\label{eq;L1}
\eea
where the ellipsis represents terms
which we do not include in our calculation.
$v^\mu$ is the velocity vector satisfying $v^2=1$;
we take $v^\mu=(1,\vec{0})$.
$S^\mu$ is the spin operator $2S^\mu=(0,\vec{\sigma})$.
$D_\mu=\partial_\mu -\frac{i}{2}\vec{\tau}\cdot\vec{\cal V}_\mu
-\frac{i}{2}{\cal V}^S_\mu$, where
$\vec{\cal V}_\mu$ and ${\cal V}_\mu^S$ are
the external isovector and isoscalar
vector currents, respectively.
$f^+_{\mu\nu} = \frac{\vec{\tau}}{2} \cdot \left(
\partial_\mu \vec{\cal V}_\nu -\partial_\nu\vec{\cal V}_\mu \right)$ and
$f^S_{\mu\nu} = \frac12\left(
\partial_\mu {\cal V}^S_\nu - \partial_\nu {\cal V}^S_\mu \right)$.
$m_N$ is the nucleon mass
and $\kappa_V$ ($\kappa_S$) is
the isovector (isoscalar)
anomalous magnetic moment of the nucleon;
$\kappa_V= 3.70589$
($\kappa_S= -0.12019$).
The terms of $\tilde{d}_7$ and $\tilde{e}_{54}$ 
correspond to those of $d_7$ and $e_{54}$ in 
the pionful one-nucleon lagrangian~\cite{fmms-ap00}
and the LEC's $\tilde{d}_7$ and $\tilde{e}_{54}$ 
are fixed by isoscalar radii of nucleon EM form factors.

${\cal L}_{s}$, ${\cal L}_t$,
and ${\cal L}_{st}$
for the dibaryon fields \cite{bs-npa01} 
may read
\bea
{\cal L}_s &=&
\sigma_s s_a^\dagger\left[iv\cdot D
+\frac{1}{4m_N}[(v\cdot D)^2-D^2]
+\Delta_s\right] s_a
-y_s\left[s_a^\dagger (N^TP^{(^1S_0)}_aN)
+ \mbox{h.c.}\right],
\label{eq;Ls}
\\
{\cal L}_t &=&
\sigma_t t_i^\dagger\left[iv\cdot D
+\frac{1}{4m_N}[(v\cdot D)^2-D^2]
+\Delta_t
\right] t_i
-y_t\left[t_i^\dagger (N^TP_i^{(^3S_1)}N) + \mbox{h.c.}\right]
\nnb \\ &&
-\frac{C_2^{(sd)}}{\sqrt{m_N\rho_d}}
{\cal T}^{(sd)}_{ij,xy}\left[
t_i^\dagger (N^T{\cal O}_{xy,j}^{(2)}N) + \mbox{h.c.}
\right]
- \frac{2L_2}{m_N\rho_d}
 i\epsilon_{ijk}t_i^\dagger t_j B_k
\nnb \\ &&
-\frac{L_2'}{\sqrt{m_N\rho_d}}
\left[ i \epsilon_{ijk}t_i^\dagger
\left(N^TP^{(^3S_1)}_jN\right) B_k
+ \mbox{h.c.}
\right]
+ \frac{2C_M}{m_N\rho_d} i \epsilon_{ijk}
t_i^\dagger \{D^2,B_j\} t_k
\nnb \\ &&
- \frac{C_Q}{m_N\rho_d} t^\dagger_i \left[
iv\cdot D, {\cal O}_{ij}^{(2)}\right] t_j ,
\label{eq;Lt}
\\
{\cal L}_{st} &=&
\frac{L_1}{m_N\sqrt{r_0\rho_d}}
\left[ t_i^\dagger s_3 B_i
+ \mbox{h.c.}\right]
+ \frac{L_1'}{\sqrt{m_N\rho_d}}\left[
t_i^\dagger \left(N^TP_3^{(^1S_0)}N\right) B_i + \mbox{h.c.}
\right]
\nnb \\ &&
+ \frac{L_1'}{\sqrt{m_Nr_0}}\left[
\left(N^TP_i^{(^3S_1)}N\right)^\dagger s_3 B_i
+ \mbox{h.c.}
\right]\, , 
\label{eq;Lst}
\eea
where the covariant derivative 
for the dibaryon field is given by
$D_\mu =\partial_\mu -iC{\cal V}^{ext}_\mu$, where
${\cal V}_\mu^{ext}$ is the external vector field.\footnote{
Since the external vector fields coupled with charges of
proton and neutron are given by $\frac12({\cal V}_\mu^S+{\cal V}^3_\mu)$
and $\frac12({\cal V}_\mu^S-{\cal V}^3_\mu)$, respectively,
${\cal V}_\mu^{exp}$ for 
$np$ channel of the dibaryon field,
for example,
is obtained by a sum of the 
charge operators of neutron and proton,
${\cal V}_\mu^{ext}(np) = \frac12 ({\cal V}_\mu^S - {\cal V}^3_\mu)
+ \frac12({\cal V}_\mu^S + {\cal V}^3_\mu) = {\cal V}^S_\mu$.
Similarly, one has
${\cal V}^{ext}_\mu(pp) = {\cal V}^S_\mu + {\cal V}^3_\mu$
and ${\cal V}^{ext}_\mu(nn) = {\cal V}^S_\mu - {\cal V}^3_\mu$.
}
$C$ is the charge operator of the dibaryon fields
and $C=0,1,2$ for the $nn$, $np$, $pp$ channel, respectively
(where we have put $e=1$).
$\vec{B}$ is the magnetic field given by
$\vec{B}=\vec{\nabla} \times\vec{\cal V}^{ext}$.
The sign factors, $\sigma_{s}$ and $\sigma_{t}$,
turn out to be $-1$ (see Appendix for details).
$\Delta_t$ ($\Delta_s$) is the difference
between the dibaryon mass $m_t$ ($m_s$)
in the $^3S_1$ ($^1S_0$) channel
and the two-nucleon mass;
$m_{t,s} = 2 m_N + \Delta_{t,s}$.
$\rho_d$ and $r_0$ are
the effective ranges for the deuteron
and $^1 S_0$ scattering state, respectively.
$P_i^{(S)}$ is the projection operator
for the $S={}^3S_1$ or $^1S_0$ channel;
\bea
&&
P_i^{({}^3S_1)} = \frac{1}{\sqrt{8}}\sigma_2\sigma_i\tau_2,
\ \ \ 
P_a^{({}^1S_0)} = \frac{1}{\sqrt{8}}\sigma_2\tau_2\tau_a,
\ \ \ 
{\rm Tr}(P_i^{(S)\dagger} P^{(S)}_j) = \frac12\delta_{ij},
\eea
where $\sigma_i$ ($\tau_a$) is the spin (isospin) operator.
The operators for the $D$-state read
\bea
{\cal T}^{(sd)}_{ij,xy} &=&
\delta_{ix}\delta_{jy} - \frac13\delta_{ij}\delta_{xy},
\ \ \
{\cal O}^{(2)}_{ij} =
- \left( D_i D_j
- \frac13\delta_{ij}\stackrel{\to}{D}^2
\right),
\\
{\cal O}^{(2)}_{xy,j} &=& -\frac14\left(
\stackrel{\leftarrow}{D}_x\stackrel{\leftarrow}{D}_yP_j^{({}^3S_1)}
+ P_j^{({}^3S_1)}\stackrel{\to}{D}_x\stackrel{\to}{D}_y
- \stackrel{\leftarrow}{D}_x P_j^{({}^3S_1)} \stackrel{\to}{D}_y
- \stackrel{\leftarrow}{D}_y P_j^{({}^3S_1)} \stackrel{\to}{D}_x
\right)\, .
\eea

The LEC's, $y_s$ and $y_t$, represent the dibaryon-$NN$
($dNN$) couplings in the spin singlet and 
triple states, respectively,
and they contribute to the two-nucleon loop 
diagram for the two-nucleon(dibaryon) propagator.
They, as well as $\Delta_{s,t}$ and $\sigma_{s,t}$,
have been determined from the
effective ranges in the ${}^1S_0$ scattering
and the deuteron states.
$C^{(sd)}_2$ accounts for $^3 D_1$ state mixture to
the $^3 S_1$ state
and is fixed by the asymptotic $D$-$S$ ratio
$\eta_{sd}$ ($\simeq$ 0.0254).
Details are given in Appendix.

We consider LEC's 
with the external vector field
in the literatures.
$C_Q$, $L_1$, and $L_2$ are the LEC's 
for the $Vdd$ vertices,
and the coefficients, $1/\sqrt{m_N\rho_d}$ and 
$1/\sqrt{m_Nr_0}$, of these terms in Eqs.~(\ref{eq;Lt}) and 
(\ref{eq;Lst}) are introduced 
by a rule in Eq.~(11) in Ref.~\cite{bs-npa01}
to convert the two-nucleon field 
($NN$) in pionless EFT 
to a dibaryon field ($t$ or $s$) in dEFT.
In this work we do not derive the most general 
effective lagrangian with dibaryons and without
pions.
$C_Q$ is determined
from the electric quadrupole moment of the deuteron,
and $L_1$ and $L_2$ 
are determined from the total cross section of the radiative
neutron capture by the proton
and the magnetic moment of the deuteron, respectively.
We introduce a new LEC $C_M$,
which can be fixed from the radius of the
magnetic form factor of the deuteron.
In general, one also has
LEC's $L_1'$ and $L_2'$ for the 
vector(photon)-dibaryon-$NN$ ($VdNN$) vertices
by using the rule mentioned above.
Since they are higher dimensional terms,
they could be considered 
to account for a higher order correction 
to the leading contributions of EM transitions.
In the next two sections,
we determine the EM LEC's
listed above and proceed
to calculate the EM observables 
in the elastic $e$-$d$ scattering.

\vskip 3mm \noindent
{\bf 3. EM form factors of the deuteron}

The EM form factors of the deuteron have been intensively
studied within the 
Weinberg's approach~\cite{pcnpa00,p-plb03},
KSW scheme~\cite{ksw-prc99},
and pionless EFT~\cite{crs-npa99}.
The electric form factor of the deuteron has also been 
calculated in dEFT without pions~\cite{bs-npa01}.
In this section we consider the
deuteron EM form factors in dEFT without pions.

A deuteron state $|\vec{p},i\rangle$
specified by momentum $\vec{p}$
and spin $i$ satisfies the normalization condition
$\langle \vec{p'},i|\vec{p},j\rangle
=(2\pi)^3\delta^{(3)}(\vec{p'}-\vec{p})\delta_{ij}$.
The nonrelativistic expansion of the matrix element of the
electromagnetic current up to 
${\cal O}(Q^3)$
is given as~\cite{crs-npa99}
\bea
\langle\vec{p'},i|J^0_{em}|\vec{p},j\rangle
&=& e\left[
F_C(q)\delta_{ij}
+\frac{1}{2m_d^2}F_Q(q)\left(q_iq_j-\frac{1}{n-1}
q^2\delta_{ij}
\right)\right]
\left(\frac{E'+E}{2m_d}\right) ,
\nnb \\
\langle\vec{p'},i|J^k_{em}|\vec{p},j\rangle
&=& \frac{e}{2m_d}\left[
F_C(q)\delta_{ij}(\vec{p'}+\vec{p})^k
+F_M(q)(\delta^k_iq_j-\delta^k_jq_i)
\right.
\nnb \\ && \left. +
\frac{1}{2m_d^2}F_Q(q)
\left(q_iq_j-\frac{1}{n-1}
q^2\delta_{ij}\right)
(\vec{p'}+\vec{p})^k
\right] ,
\label{eq;FFs}
\eea
where $\vec{p'}=\vec{p}+\vec{q}$
and $q=|\vec{q}|$.
$e$ is the electric charge,
$m_d$ the deuteron mass,
and $E$ ($E'$) the energy of the deuteron
in the initial (final) state.
$n$ is the space-time dimensions, $n=4$.
$F_C(q)$, $F_M(q)$, and $F_Q(q)$
represent electric charge, 
magnetic dipole, and electric quadrupole form factors,
respectively. 
These dimensionless form factors
defined in Eq. (\ref{eq;FFs})
are conventionally normalized as
\bea
F_C(0) = 1,
\ \ \
\frac{e}{2m_d}F_M(0) = \mu_M,
\ \ \
\frac{1}{m_d^2}F_Q(0) = \mu_Q,
\label{eq;FFnorm}
\eea
where $\mu_M (\equiv \mu_d\, \frac{e}{2m_N})$
is the magnetic moment of the deuteron
and $\mu_Q$ is its electric quadrupole moment:
$\mu_d = 0.8574382284 (94)$~\cite{mt-rmp00} 
and $\mu_Q = 0.2859 (3)$ fm$^2$~\cite{er-npa83}.
The charge radius of the deuteron
$\sqrt{\langle r_{ch}^2\rangle}$ is defined by
$F_C(q) = 1-\frac16\langle r_{ch}^2\rangle q^2 +\cdots$
and its empirical value is $\sqrt{\langle r_{ch}^2\rangle} =
2.1303 (10)$ fm~\cite{st-npa98}.

\vskip 3mm \noindent
{\bf 3.1 Electric form factor}

\begin{figure}[tbp]
\begin{center}
\epsfig{file=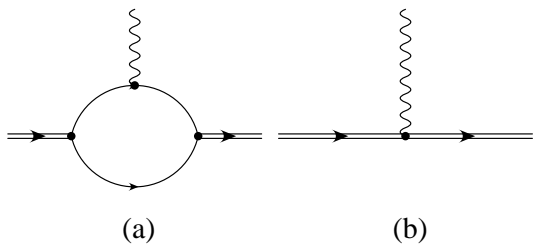,width=7cm}
\caption{\label{fig;FC-d}
Diagrams for deuteron electric form factor $F_C(q)$
at LO, ${\cal O}(Q^0)$.
The single line stands for the nucleon field, 
the double line for the dibaryon,
and the wavy line for the external vector field.
The vertex with a dot coupled 
with the external vector field in (a) is proportional to $v^\mu$, 
while that with a dot in (b) is proportional to $v^\mu \sigma_t$.  
The $dNN$ vertex with a dot in (a) is 
proportional to $y_t$.
}
\end{center}
\end{figure}

Two LO, ${\cal O}(Q^0)$, diagrams 
in Fig.~\ref{fig;FC-d} contribute to 
the electric form factor of the deuteron, $F_C(q)$.
Three-point vertex functions calculated from 
Fig.~\ref{fig;FC-d} (a) and (b) read
\bea
i\Gamma^{(a)\mu}_{(ij)} &=&
-i\delta_{ij} v^\mu\sigma_t \frac{4}{\rho_d q}
{\rm arctan}\left(\frac{q}{4\gamma}\right),
\\
i \Gamma^{(b)\mu}_{(ij)} &=& i \delta_{ij}v^\mu \sigma_t ,
\eea
where $i$ and $j$ denote the final and the initial spin state of
the deuteron, respectively;
$\mu$ is the Lorentz index for the current.
Multiplying the three-point vertices
with the normalization factor $Z_d$ of the deuteron wave function
($Z_d$ is derived in Eq.~(\ref{eq;Zd}) in Appendix),
one obtains the charge form factor of the deuteron at LO as
\bea
v^\mu F^{LO}_C(q) \delta_{ij}
= Z_d\left[
\Gamma^{(a)\mu}_{(ij)}+\Gamma^{(b)\mu}_{(ij)}
\right],
\eea
and thus
\bea
F^{LO}_C(q) = \frac{\gamma\rho_d}{1-\gamma\rho_d}
\left[\frac{4}{\rho_d q}
{\rm arctan}\left(\frac{q}{4\gamma}\right)
-1
\right] .
\label{eq;FC}
\eea
This result is 
the same as that 
calculated with ERT \cite{ksw-prc99}.  
We obtain 
the same charge radius $\langle r_{ch}^2\rangle$
as ERT,
$\langle r_{ch}^2\rangle^{ER} = \frac{1}{8\gamma^2}
\frac{1}{1-\gamma\rho_d}$ = (1.985)$^2$ fm$^2$.

Higher order corrections 
to $F_C(q)$ have been studied,
{\it e.g.}, in pionless EFT~\cite{crs-npa99},
but a slightly different counting rule 
is employed in it where
one expands 
the form factors 
in terms of the effective range,
{\it i.e.}, counts $\gamma\sim Q$, $1/\rho_d\sim Q^0$, 
and $\gamma\rho_d\sim Q$.
(See also footnote \ref{footnote;counting}.)
After expanding our result in terms of ($\gamma\rho_d)$,
one can easily compare 
it with Eqs.~(3.21,22,23) in Ref.~\cite{crs-npa99} 
and find agreement with each other
up to the subleading order.
A difference in the higher order 
comes out of
the isoscalar vector radius of nucleon current
and relativistic correction.

One can verify that
$F_C^{LO}(q)$ in Eq.~(\ref{eq;FC}) satisfies
the normalization condition given by Eq.~(\ref{eq;FFnorm}).
It would be worth noting that 
the expression of $F_C^{LO}(q)$ 
has no free parameter
because the leading $Vdd$ vertex that
contributes to the electric form factor stems from
the covariant derivative of the dibaryon field
(Fig.~\ref{fig;FC-d} (b)),
which is proportional to
the dibaryon charge $C=1$ for the $np$ channel 
and the overall factor $\sigma_t$.
In other words, the magnitude of strength of the $Vdd$ vertex
is determined solely by the 
charges of nucleons
associated with the one-body $VNN$ vertex. 
At $q = 0$,
the contribution of the diagram
with $Vdd$ vertex (Fig.~\ref{fig;FC-d} (b))
amounts to $-$40\% of that with 
the $VNN$ vertex (Fig.~\ref{fig;FC-d} (a)).
As will be shown in the next subsections,
we find similar values of the $Vdd$ to $VNN$ ratios
for the other form factors.
More importantly,
the quantity in the bracket of Eq.~(\ref{eq;FC})
exactly cancels the factor 
$\gamma\rho_d/(1-\gamma\rho_d)$,
the deuteron normalization factor $Z_d$,
at $q = 0$, 
which leads us to assign a new
role to the LEC's of the $Vdd$ vertices.

\vskip 3mm \noindent
{\bf 3.2 Magnetic form factor}

\begin{figure}[tbp]
\begin{center}
\epsfig{file=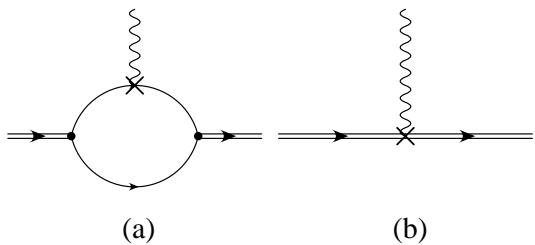,width=7cm}
\caption{\label{fig;FM-d}
Diagrams contributing to 
deuteron magnetic form factor 
$F_M(q)$.
The leading order diagram (a) is of 
${\cal O}(Q)$,
while the next-to-leading order diagram (b) is 
of $\calO(Q^2)$.
The vertex with ``X'' in (a) is proportional to
$(1+\kappa_S)/(2m_N)$ while that with ``X'' in (b)
is proportional to $L_2$.
See the caption for Fig.~\ref{fig;FC-d} as well.
}
\end{center}
\end{figure}

Diagrams contributing to the magnetic form factor
of the deuteron are depicted in Fig. \ref{fig;FM-d}.
The order of Fig.~\ref{fig;FM-d} (a) is LO ($\calO(Q)$)
and that of Fig.~\ref{fig;FM-d} (b) 
is next-to-leading order (NLO) ($\calO(Q^2)$)
for the vertex functions.
We will discuss however that
the contribution of Fig.~\ref{fig;FM-d} (b)
can be separated into LO and sub-leading parts, and show that
this re-ordering 
makes the convergence more efficient.
The three-point vertex for each diagram is 
\bea
i\Gamma^{(a)}_{(ij)k} &=&
i(\delta_{ik}q_j-\delta_{jk}q_i)
\frac{1+\kappa_S}{2m_N}\frac{4}{\rho_d q}
{\rm arctan}\left(\frac{q}{4\gamma}\right),
\\
i\Gamma_{(ij)k}^{(b)} &=&
i (\delta_{ik}q_j-\delta_{jk}q_i)
\frac{2L_2}{m_N\rho_d} ,
\label{eq;GMb}
\eea
where $k$ is the index
for the spatial part of the vector current.
Multiplying the three-point functions with
the normalization factor $Z_d$,
one has the magnetic form factor $F_M(q)$,
\bea
\frac{1}{2m_d}F_M(q)
= \frac{\gamma\rho_d}{1-\gamma\rho_d} \left[
\frac{1+\kappa_S}{2m_N} \frac{4}{\rho_d q}
{\rm arctan}\left(\frac{q}{4\gamma}\right)
+\frac{2L_2}{m_N\rho_d}
\right].
\label{eq;FM}
\eea
At $q=0$,
\bea
\frac{1}{2m_d}F_M(0)
= \frac{\gamma\rho_d}{1-\gamma\rho_d}
\left[
\frac{1+\kappa_S}{2m_N}\frac{1}{\gamma\rho_d}
+\frac{2L_2}{m_N\rho_d}
\right] \, .
\label{eq;FM0}
\eea

If one fixes the LEC $L_2$ using the 
experimental value of the deuteron magnetic moment $\mu_M$
in Eq.~(\ref{eq;FFnorm}),
one obtains $L_2 = -0.4033$ fm.
Comparing the magnitude of Fig.~\ref{fig;FM-d} (a) with
that of
Fig.~\ref{fig;FM-d} (b),
one finds that the latter is about $-$42\% of the former. 
As mentioned in the previous subsection, this ratio
is similar to the $Vdd$ to $VNN$ ratio
in $F_C^{LO}(0)$.
Since the former EFT calculations \cite{crs-npa99,pkmr-prc98}
show that the LO one-body term is dominant in 
the estimations of $\mu_M$
and the higher order 
terms give only a few percents correction,
this rather large ratio we obtained here 
causes a difficulty to interpret $L_2$ in dEFT
as the high energy contributions integrated out from the 
effective lagrangian.
So we may divide $L_2$ 
into the sum of the LO and sub-leading contributions.\footnote{
The same partition
has been employed for the LEC $L_2$
in the recent dEFT calculations\cite{ds-04,cjl-04}.
}
This modified counting of the LEC $L_2$ reads
\bea
L_2 = L_2^0 + \delta L_2 , \ \ \
L_2^{0} = -\frac14(1+\kappa_S)\rho_d,
\label{eq;L20}
\eea
where $L^0_2$ is LO, while $\delta L_2$ accounts
for the sub-leading contributions.
Inserting $L^0_2$ into $L_2$ of the lagrangian,
one finds that the coefficient of the $Vdd$ interaction
becomes $(1 + \kappa_S)/(2 m_N)$
which is the same as that of the one-body interaction.
Note that $L^0_2$ is of $Q^{-1}$
because of the factor $\rho_d$,
so the diagram Fig.~\ref{fig;FM-d} (b)
with $L^0_2$
becomes the same order as
the diagram Fig.~\ref{fig;FM-d} (a).
With the modified counting of $L_2$
the magnetic form factor at LO becomes
\begin{eqnarray}
\left. \frac{1}{2 m_d} F_M (q) \right|_{L_2=L_2^0}
&=& \frac{1 + \kappa_S}{2 m_N}\,
\frac{\gamma \rho_d}{1 - \gamma \rho_d}
\left[\frac{4}{\rho_d q}\arctan\left(\frac{q}{4 \gamma}\right) -1 \right]
\nonumber \\ &=&
\frac{1 + \kappa_S}{2 m_N}\, F^{LO}_C(q),
\end{eqnarray}
which is equivalent to the relation of 
$F_M(q)$ and $F_C(q)$ obtained from
ERT \cite{ksw-prc99}.
Numerically, one finds $L^0_2 \simeq -0.3880$ fm.
This value is
larger than $L_2$ fixed from Eq.~(\ref{eq;FFnorm}) by about 3.7 \%.
This 3.7 \% difference can be attributed to the 
contribution of $\delta L_2$ 
(or the higher order term of $L_2'$. 
We will discuss it below.) 
It is also easy to compare our LO result expanding 
in terms of $\gamma\rho_d$ with those of 
pionless EFT calculation, 
Eqs.~(3.32,33) in Ref.~\cite{crs-npa99}. 
We find good agreement with each other
and that 
our $\delta L_2$ plays a similar role to the 
LEC in pionless EFT.
Thus the partition
of LEC $L_2$ in Eq.~(\ref{eq;L20}) 
is essential to find a relation between 
EFT's with and without dibaryon fields.
\begin{figure}[tbp]
\begin{center}
\epsfig{file=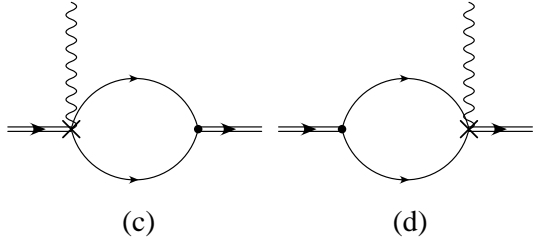,width=7cm}
\caption{\label{fig;FM-d-L2p}
Diagrams contributing 
to the deuteron magnetic form factor $F_M(q)$
arising from the $L_2'$ term.
The ``X'' in the diagrams denotes 
the $VdNN$ vertex corresponding to $L_2'$.
Both diagrams belong to ${\cal O}(Q^3)$.
}
\end{center}
\end{figure}

\begin{figure}[tbp]
\begin{center}
\epsfig{file=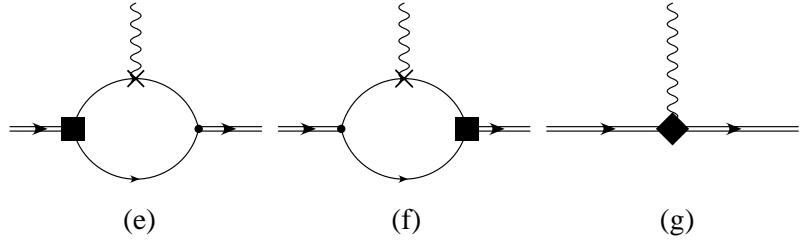,width=10.5cm}
\caption{\label{fig;FM-d-d}
Diagrams contributing
to the deuteron magnetic form factor $F_M(q)$
arising from the deuteron $D$-wave.
The diagrams (e) and (f) are 
of ${\cal O}(Q^3)$ and the diagram (g)
is of ${\cal O}(Q^4)$,
which will be considered later in $e$-$d$ scattering.
The ``X'' denotes the isoscalar magnetic vertex
of nucleon-photon coupling
proportional to $(1+\kappa_S)/(2m_N)$,
and the filled boxes stand for 
the deuteron $D$-wave in the diagrams (e) and (f).
The filled diamond in the diagram (g) is
a vertex proportional to $C_M$.
}
\end{center}
\end{figure}

Next we consider some of 
higher order contributions
from $L_2'$ of the $VdNN$ vertex (Fig.~\ref{fig;FM-d-L2p})
and the $D$-wave of the deuteron (Fig.~\ref{fig;FM-d-d}).
Since there are no corrections 
to the isoscalar magnetic $VNN$ vertex
from the third order heavy-baryon
chiral lagrangian~\cite{fmms-ap00},
we do not have
NLO corrections to $F_M(q)$
from the one-body sector.\footnote{
At NNLO, there are additional corrections
from the radius and relativistic corrections
to the isoscalar magnetic nucleon current.
Since these corrections from one-body nucleon form factors are
well known in the calculations of heavy-baryon chiral perturbation
theory (see, e.g., Ref~\cite{bfhm-npa98}), 
we will include the nucleon radii
in a later subsection
where the elastic $e$-$d$ scattering is considered.}
From the diagrams in Fig.~\ref{fig;FM-d-L2p}
one obtains a three-point vertex function
\bea
i\Gamma_{(ij)k}^{(c+d)} &=&
i(\delta_{ik}q_j-\delta_{jk}q_i)
\frac{2\gamma}{m_N \rho_d} \sqrt{\frac{m_N}{2\pi}} L_2',
\eea
where dimensional regularization 
has been used for the loop integration.
We find 
that contributions from the LEC's $L_2'$ and $\delta L_2$
have the same momentum dependence in $F_M(q)$ 
and thus the two LEC's are redundant constants.
In this work, we choose $\delta L_2=0$ 
and $L_2'$ is fixed 
by the deuteron magnetic moment below.
%
Sum of the diagrams in Fig.~\ref{fig;FM-d-d} gives
\bea
i\Gamma_{(ij)k}^{(e+f)} &=&
i(\delta_{ik}q_j-\delta_{jk}q_i)
\frac{\sqrt{2}(1+\kappa_S)\eta_{sd}}{32m_N\rho_d\gamma^2q}
\left[
4\gamma q - (16\gamma^2+3q^2)
{\rm arctan}\left(\frac{q}{4\gamma}\right)
\right] ,
\nnb \\
i\Gamma_{(ij)k}^{(g)} &=&
i(\delta_{ik}q_j-\delta_{jk}q_i)
\frac{2C_M}{m_N\rho_d} (\vec{p'}^2+\vec{p}^2)\, .
\label{eq;GMefg}
\eea

Summing up the contributions from Fig.~\ref{fig;FM-d} to Fig.~4,
we obtain the magnetic form factor of the deuteron 
with the NNLO corrections 
considered here as
\bea
\frac{1}{2m_d}F_M(q) &=&
\frac{\gamma\rho_d}{1-\gamma\rho_d}
\left\{
\frac{1+\kappa_S}{2m_N}\left[
\frac{4}{\rho_dq}{\rm arctan}\left(\frac{q}{4\gamma}\right)
-1
\right.\right.
\nnb \\ && \left.\left.
+\frac{\sqrt{2}\eta_{sd}}{16\gamma^2\rho_dq}\left(
4\gamma q
-(16\gamma^2+3q^2){\rm arctan}\left(\frac{q}{4\gamma}\right)
\right)\right]
\right. \nnb \\ && \left.
+\frac{2\gamma}{m_N\rho_d}\sqrt{\frac{m_N}{2\pi}} L_2'
\right\} \, , 
\eea
where we have put $\delta L_2 = 0$
({\it i.e.}, $L_2 = L_2^0$), as mentioned before.
Fixing $L'_2$ from Eq.~(\ref{eq;FFnorm}) 
at $q=0$, 
we obtain $L_2' = -0.07096$ fm$^{5/2}$,
which gives about 2.5\% correction to the leading term. 
We will study below %
a role of the higher order corrections
to the form factor $F_M(q)$ 
in elastic $e$-$d$ scattering.

\vskip 3mm \noindent
{\bf 3.3 Electric quadrupole form factor}

\begin{figure}[tbp]
\begin{center}
\epsfig{file=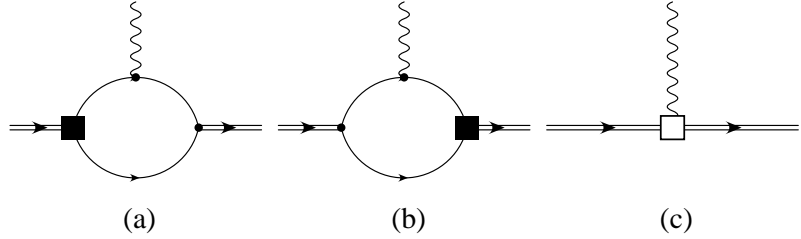,width=10.5cm}
\caption{\label{fig;FQ-d}
Diagrams for deuteron electric quadrupole
form factor $F_Q(q)$. 
Order of the diagrams (a,b) is of $\calO(Q^2)$
and that of the diagram (c) is 
of $\calO(Q^3)$.
The filled boxes in the diagrams (a,b) denote
the $D$-wave contribution and 
the empty box in the diagram (c)
represents a vertex proportional to $C_Q$.
}
\end{center}
\end{figure}

Diagrams for the electric quadrupole form factor
of the deuteron are depicted in Fig.~\ref{fig;FQ-d}.
The diagrams (a) and (b) 
in Fig.~\ref{fig;FQ-d} 
are of $\calO(Q^2)$,
while the diagram (c) in Fig.~\ref{fig;FQ-d} 
is one order higher, of $\calO(Q^3)$.
As before, 
we will see the similar $VNN$ and $Vdd$ ratio in
$F_Q(q)$ to that found in $F_M(q)$.
Thus we employ a similar assumption to
that is used for $L_2$ in Eq.~(\ref{eq;L20}) 
to extract the LO part of LEC $C_Q$
in the $Vdd$ vertex in Fig.~\ref{fig;FQ-d} (c).

Three-point vertex functions
from the diagrams in Fig.~\ref{fig;FQ-d}
are obtained as
\bea
i\Gamma^{(a+b)\mu}_{(ij)} &=&
iv^\mu \left(q_iq_j-\frac13\delta_{ij}q^2\right)
\frac{3}{4\sqrt{2}}
\frac{\eta_{sd}}{\rho_d\gamma^2q^3}
\left[
-4\gamma q
+(16\gamma^2+3q^2){\rm arctan}\left(\frac{q}{4\gamma}\right)
\right] ,
\\
i\Gamma^{(c)\mu}_{(ij)} &=& iv^\mu
\left(q_iq_j-\frac13q^2\delta_{ij}\right)
\frac{C_Q}{m_N\rho_d}\, .
\eea
Multiplying the three-point functions with the factor $Z_d$,
we obtain
\bea
\frac{F_Q(q)}{2m_d^2} &=&
\frac{\rho_d\gamma}{1-\rho_d\gamma} \left\{
\frac{3}{4\sqrt{2}}\frac{\eta_{sd}}{\rho_d\gamma^2q^3}
\left[
-4\gamma q
+(16\gamma^2+3q^2){\rm arctan}\left(\frac{q}{4\gamma}\right)
\right]
+\frac{C_Q}{m_N\rho_d}
\right\},
\label{eq;FQ}
\eea
and at $q=0$
\bea
\frac{F_Q(0)}{m_d^2} = \frac{2\rho_d\gamma}{1-\rho_d\gamma}
\left[
\frac{1}{2\sqrt{2}}\frac{\eta_{sd}}{\rho_d\gamma^3}
+ \frac{C_Q}{m_N\rho_d}
\right] \, .
\label{eq;FQq=0}
\eea

The normalization condition Eq.~(\ref{eq;FFnorm}) 
leads to $C_Q = - 1.709$ fm$^2$.
Comparing the first and second terms
in the bracket in Eq.~(\ref{eq;FQq=0}),
one finds that the latter is about $-$50\%
of the former.
In addition, it shows a pattern similar to what
we have 
seen in the cases of $F_C(q)$ and $F_M(q)$.
One may notice, however, that 
the contribution to $F_Q(q)$ is generated 
by the $D$ wave in the diagrams (a) and (b) in Fig.~\ref{fig;FQ-d}
and there is no one-body $VNN$ vertex 
corresponding to the 
$Vdd$ vertex of the LEC $C_Q$. 
We assume here that the prime portion of the LEC $C_Q$ in dEFT,
denoted by $C_Q^0$, and the contributions from the two-nucleon loop
diagrams (Fig.~\ref{fig;FQ-d} (a) and (b)) 
are combined and 
produce a factor $(1 - \rho_d \gamma)$ 
so that it cancels the factor $(1 - \rho_d \gamma)^{-1}$ in $Z_d$.
We choose a modified counting of $C_Q$ as
\bea
C_Q = C_Q^0  +\delta C_Q ,
\ \ \
C_Q^0 = -\frac{1}{2\sqrt{2}}\frac{\eta_{sd}}{\gamma^2}m_N\rho_d ,
\label{eq;CQ0}
\eea
where numerically $C_Q^0=-1.408$ fm$^2$.
The contribution of $\delta C_Q$ to $\mu_Q$,
which can be attributed to the contribution of the 
high energy physics,
is about $-$17.2\%.
This rearrangement of $C_Q$
leads to more efficient convergence in dEFT calculation.

Expanding 
Eq.~(\ref{eq;FQ})
in terms of $(\gamma\rho_d)$, 
we compare it with 
a result of pionless EFT,
Eq.~(29) in Ref.~\cite{crs-plb99}.\footnote{
Two expressions of $F_Q(q)$ have 
been reported by the same authors
in former pionless EFT calculations,
Refs.~\cite{crs-npa99} and \cite{crs-plb99},
but these two calculations showed a slightly different 
result with each other.
In Ref.~\cite{crs-npa99} 
a LEC corresponding to $C_2^{(sd)}$ was fixed 
by using a mixing parameter $\bar{\epsilon}_1$ 
in the $^3S_1$-$^3D_1$ channel 
$NN$ scattering,
while in Ref.~\cite{crs-plb99} 
it was fixed by using the asymptotic $D$-$S$ ratio $\eta_{sd}$, 
which is the same quantity as what we have used 
for fixing the LEC $C_2^{(sd)}$.
At $q=0$, 
consequently, 
numerical results of 
leading order contributions 
to the deuteron quadrupole moment, $\mu_Q^{LO}$, 
are different and 
obtained as $\mu_Q^{LO}= 0.273$ and 0.335 
fm$^2$ in Refs.~\cite{crs-npa99} and 
\cite{crs-plb99}, respectively.
Moreover, 
a subleading order correction from $\mu_Q^{LO}$ 
was obtained as $\frac12 (\gamma\rho_d)\mu_Q^{LO}$
in Eq.~(3.42) in Ref.~\cite{crs-npa99}.
The factor 1/2 of it cannot be obtained by 
expanding 
a factor $1/(1-\gamma\rho_d)$ in $Z_d$
in terms of $(\gamma\rho_d)$
and appeared 
as another difference from the result in Ref.~\cite{crs-plb99}.
Therefore, we compare our result with that in Ref.~\cite{crs-plb99}.
}
Up to next-to leading order in the 
$(\gamma\rho_d)$ expansion,
we find good agreement with each other. 
In a higher order, a difference comes from 
nucleon radius terms which are
higher order ones in our counting rule.
This agreement here, however, is found 
without introducing the partition of the LEC $C_Q$ 
in Eq.~(\ref{eq;CQ0}).
Since, as mentioned before, 
there is no one-body $VNN$ vertex corresponds to
the $Vdd$ vertex of $C_Q$, 
it may not be essential to introduce the partition 
in Eq.~(\ref{eq;CQ0}) for the LEC $C_Q$. 
%

\vskip 3mm \noindent
{\bf 3.4 Elastic electron-deuteron scattering}

With the deuteron form factors obtained above
we now make a brief study of
elastic electron-deuteron scattering, $e+d\to e+d$.
The differential cross section of the reaction is given by
\bea
\frac{d\sigma}{d\Omega}
&=&
\left.\frac{d\sigma}{d\Omega}\right|_{Mott}
\left[A(q)+B(q){\rm tan}^2\left(\frac{\theta}{2}\right)
\right].
\eea
The form factors $A(q)$ and $B(q)$ 
are related to the deuteron form factors via
\bea
A(q) = F_C^2(q) + \frac23 \eta F_M^2(q) + \frac89 \eta^2 F_Q^2(q),
\ \ \
B(q) = \frac43\eta(1+\eta)F_M^2(q),
\label{eq;AB}
\eea
with $\eta=q^2/(4m_d^2)$.

In incorporating 
effect of the nucleon radius into
the electric form factor $F_C(q)$,
which have not been considered explicitly
in the previous subsections,
we introduce the radius of the isoscalar vector nucleon current
and a radius of the dibaryon field
into the diagrams (a) and (b) in Fig.~\ref{fig;FC-d},
respectively.
As discussed so far,
order of a LEC of $Vdd$ vertex, here for the dibaryon radius, 
should be modified and parted into 
LO and subleading part and  
the large potion of the LEC can be fixed by the nucleon radius.
From the large portion of the LEC and the nucleon radius, 
in consequence,
one has an expression as introducing 
the nucleon form factor (where we retain only the radius term)
into $F_C^{LO}(q)$ in Eq.~(\ref{eq;FC}).
From the small portion of the LEC 
one has a parameter to fix. 
Thus one has
\bea
F_C(q) = \frac{\rho_d \gamma}{1-\rho_d\gamma}
\left\{
\left[\frac{4}{\rho_d q}{\rm arctan}\left(\frac{q}{4\gamma}\right)
-1\right]\left(1-\frac16\langle r_{Es}^2\rangle q^2\right)
-\frac16 \delta_c q^2 \right\}\, ,
\eea
where $\langle r_{Es}^2\rangle$ 
is the isoscalar-electric radius
of the nucleon Sachs form factor,\footnote{
We have obtained the values of the nucleon radii
$\langle r_{Es}^2\rangle$
and $\langle r_{Ms}^2\rangle$
from $r_1^{(s)}$ and $r_2^{(s)}$, the isoscalar
radii for the Dirac and Pauli form factors,
in Table 1 of Ref.~\cite{metal-npa96}.  }
$\langle r_{Es}^2\rangle=0.604$ fm$^2$,
and relates to the LEC $\tilde{d}_7$ in
the one-nucleon lagrangian ${\cal L}_N$ in 
Eq.~(\ref{eq;L1}) via
$\langle r_{Es}^2\rangle=-6\tilde{d}_7$. 
$\delta_c$ is a subleading radius correction
from the dibaryon field. 
The corrected charge radius reads
$\langle r_{ch}^2\rangle = 
\frac{1}{1-\rho_d\gamma} 
\frac{1}{8\gamma^2}
+ \langle r_{Es}^2\rangle
+ Z_d\delta_c$.
Using the experimental value
of $\langle r_{ch}^2\rangle$,
we obtain $\delta_c=-0.011$ fm$^2$.
Similarly, introduction of the nucleon radius
in $F_M(q)$ leads to 
\bea
\frac{e}{2m_d} F_M(q) &=&
\frac{e\gamma\rho_d}{1-\gamma\rho_d} \left\{
\frac{1+\kappa_S}{2m_N}\left(1-\frac16\langle r_{Ms}^2\rangle q^2\right)
\left[
\frac{4}{\rho_dq}{\rm arctan}\left(\frac{q}{4\gamma}\right)
-1
\right.\right. \nnb \\ &&
\left.
+\frac{\sqrt{2}\eta_{sd}}{16\gamma^2 \rho_d q}\left(
4\gamma q-(16\gamma^2+3q^2){\rm arctan}\left(\frac{q}{4\gamma}\right)
\right) \right]
\nnb \\ && \left.
+\frac{2\gamma}{m_N\rho_d} \sqrt{\frac{m_N}{2\pi}} L_2'
+ \frac{2\delta C_M}{m_N\rho_d}q^2
\right\}\, ,
\label{eq;rV2}
\eea
where we have separated the LEC $C_M=C_M^0+\delta C_M$
(and one has
$C_M^0=(1+\kappa_S)\rho_d\langle r_{Ms}^2\rangle/24$)
and
fixed the value of $C_M^0$ by the
one-body radius, isoscalar-magnetic radius of
the Sachs form factor of the nucleon,
$\langle r_{Ms}^2\rangle 
= 0.598$ fm$^2$. 
In Eq.~(\ref{eq;rV2})
we also have multiplied the 
$D$-wave corrections 
from the diagrams (e) and (f) in Fig.~\ref{fig;FM-d-d}
with the nucleon form factor 
because the photon couples
directly with the nucleon in these diagrams.
Since both of the corrections from the $D$ waves and nucleon radius
are of NNLO,
this multiplication affects to terms in N$^4$LO.
$\langle r_{Ms}^2\rangle$ 
relates to the LEC $\tilde{e}_{54}$ in the one-nucleon 
lagrangian ${\cal L}_N$ in Eq.~(\ref{eq;L1})
via
$\langle r_{Ms}^2\rangle 
=24m_N\tilde{e}_{54}/(1+\kappa_S)$.

LEC $\delta C_M$ can be fixed from the 
magnetic radius of the deuteron,
$\sqrt{\langle r_{Md}^2\rangle}$ of $F_M(q)$. 
From Eq. (\ref{eq;rV2}) one obtains the relation 
for $\langle r_{Md}^2\rangle$,
\bea
\langle r_{Md}^2 \rangle &=&
\frac{e}{\mu_M}\left\{
\frac{1+\kappa_S}{2m_N}\left[
\frac{1+\sqrt{8}\eta_{sd}}{(1-\gamma\rho_d)8\gamma^2}
+ \langle r_{Ms}^2\rangle
\right]
- \frac{12\gamma \delta C_M}{(1-\gamma\rho_d)m_N}
\right\} \, .
\eea
Precise empirical value of $\langle r^2_{Md} \rangle$ is not
available yet, so we choose a value of 
$\delta C_M$, equivalently,
a value of 
$\sqrt{\langle r^2_{Md}} \rangle$, 
as $\sqrt{\langle r^2_{Md}} \rangle
= 2.135$ fm,
so as to get a better fitting for
experimental data of $B(q)$ \cite{ssw-npa81} 
up to momentum transfer 400 MeV. 
We ignore the radius effects in
$F_Q(q)$ because
the contribution of $F_Q(q)$ to $A(q)$ is negligible
due to the factor $\eta^2$ (see Eq.~(\ref{eq;AB}))
in the momentum range under consideration. 

\begin{figure}[tbp]
\begin{center}
\epsfig{file=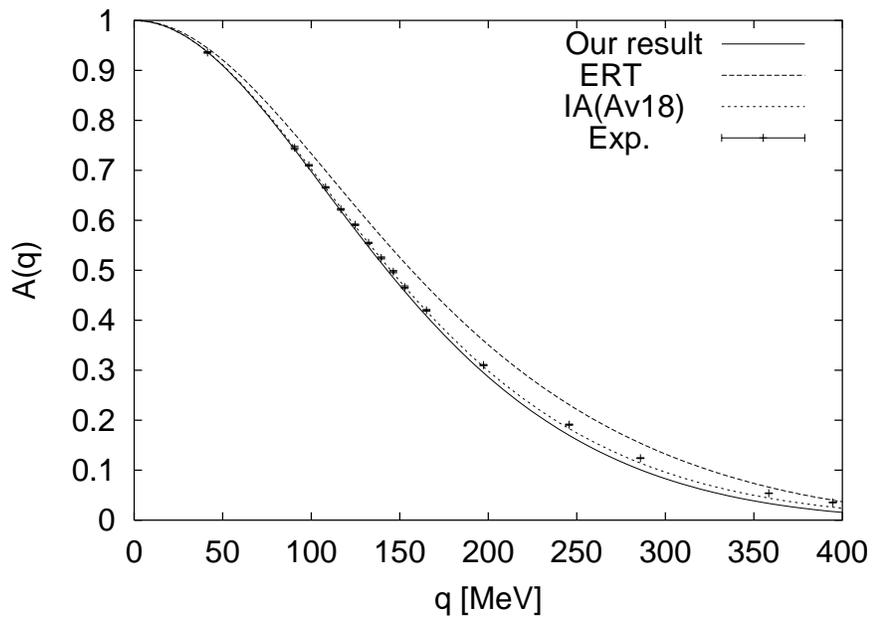,width=12cm}
\caption{\label{fig;A}
Form factor $A(q)$ in the elastic $e$-$d$ scattering.
The solid line corresponds to our result,
the dashed line to the result of ERT,
and the dotted line to that of 
the potential model calculation 
that employs the one-body (IA) operator~\cite{j-pr56},
the deuteron wave functions obtained from the Argonne v18 potential,
and the nucleon radii from Ref.~\cite{metal-npa96}.
The experimental data are taken from Ref.~\cite{ssw-npa81}.
}
\end{center}
\end{figure}
\begin{figure}[tbp]
\begin{center}
\epsfig{file=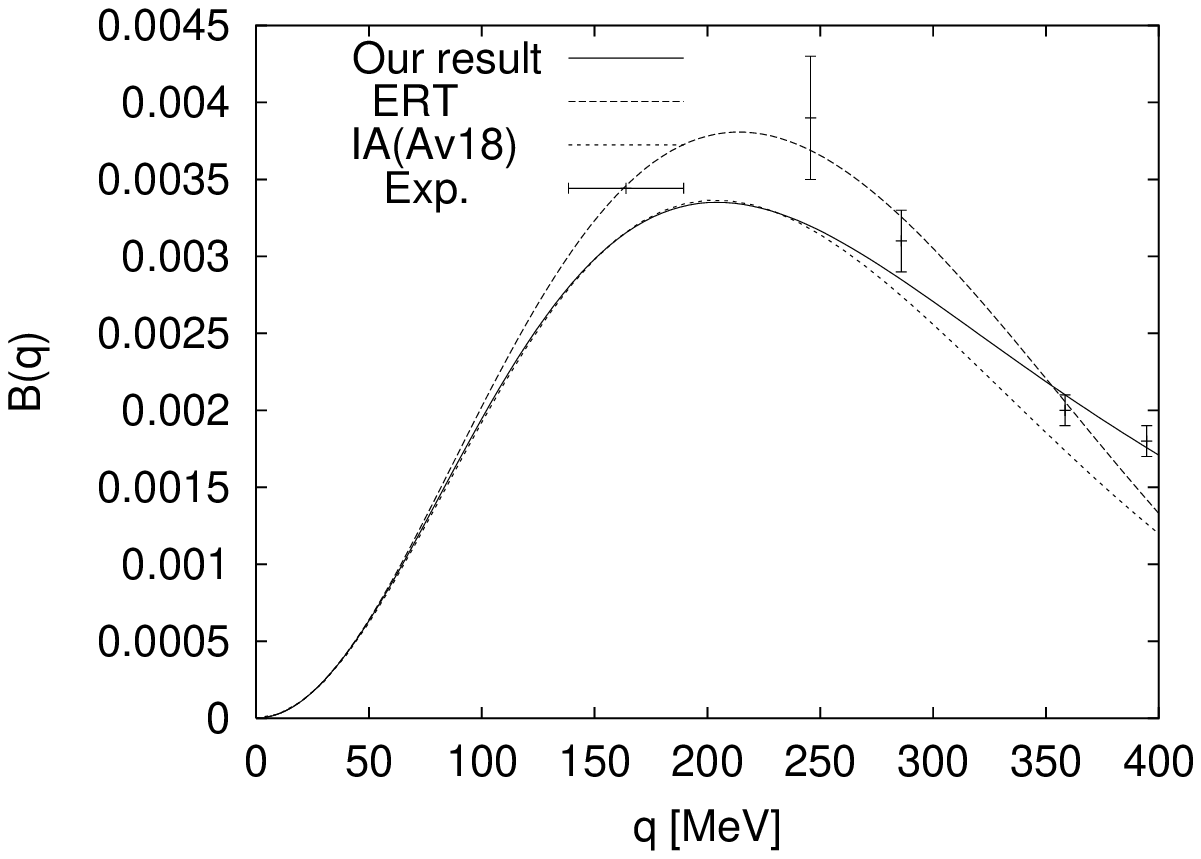,width=12cm}
\caption{\label{fig;B}
Form factor $B(q)$ in the elastic $e$-$d$ scattering.
For further explanations, 
see the caption 
of Fig.~\ref{fig;A}.
}
\end{center}
\end{figure}
The results for $A(q)$ and $B(q)$ 
are plotted up to $q=400$ MeV
in Figs.~\ref{fig;A} and \ref{fig;B}, respectively.
Since our formalism is a pionless theory,
the momentum transfers
up to $q=400$ MeV may be too large 
(compared to the pion mass) and it could not be 
suitable for the theory without pions. 
However, we show the higher momentum transfer
region to compare our result with 
the existing low-energy experimental data of $B(q)$.
In both figures,
the solid lines represent our present results,
the dashed lines correspond to the results of ERT,
and the dotted lines to those of 
a potential model calculation.
In the potential model calculation, 
we employed 
the one-body (IA) current operator\cite{j-pr56},
the deuteron wave functions obtained 
from the Argonne v18 potential (Av18),
and the nucleon radii from Ref.~\cite{metal-npa96}.
The experimental data 
are taken from Ref.~\cite{ssw-npa81}.
Our LO dEFT calculations reproduce well
the results of ERT in the small momentum region.
With the higher order corrections included,
the results of the dEFT approach 
become better than those of ERT and are 
comparable to those of the accurate potential model calculations
for both $A(q)$ and $B(q)$ 
up to about $q=200$ MeV.
Values of $r_{ch}$, $\mu_d$ and $r_{Md}$ 
calculated from 
ERT and Av18 are compared with the ones used 
in our calculation in Table~\ref{tab;compare}.
Experimental values of them can be found
{\it e.g.} in Table 1 of Ref.~\cite{go-anp01}.
We should note that the experimental value
of $r_{Md}$ still has relatively large uncertainty
compared to the accurate empirical value of $r_{ch}$.
More accurate measurements of $B(q)$ 
are expected to reduce the uncertainty of $r_{Md}$.
\begin{table}[t]
\begin{center}
\begin{tabular}{ c | c c c c }\hline
          & ERT   & IA(Av18) & dEFT      & Exp.\cite{go-anp01} \\ \hline
$r_{ch}$  & 1.985 & 2.115    & 2.130$^*$ & 2.130 \\ 
$r_{Md}$  & 1.985 & 2.090    & 2.135$^*$ & $--$  \\ 
$\mu_d$   & 0.880 & 0.847    & 0.857$^*$ & 0.857 \\ \hline
\end{tabular}
\end{center}
\caption{
Comparison of the physical observables; radii,
$r_{ch}$ and $r_{Md}$, of the electric and magnetic form factors
and the magnetic moment $\mu_d$ of the deuteron.
$r_{ch}$ and $r_{Md}$ are given in units of fm.
In the column representing dEFT, 
the entries with '$*$' are input
used for fixing the LEC's. 
Experimental values can be found, 
{\it e.g.}, in Table 1 of Ref.~\cite{go-anp01}.
}
\label{tab;compare}
\end{table}

\vskip 3mm \noindent
{\bf 4. Radiative neutron capture by the proton }

Radiative neutron capture by the proton,
$n+p\to d+\gamma$, has been intensively studied
within the frameworks of EFT
employing the Weinberg's scheme~\cite{pmr-prl95} and
the KSW scheme~\cite{ssw-npa99}
and pionless EFT~\cite{crs-npa99};
see also~\cite{bs-npa01}.
We study this reaction 
in dEFT in more detail here.

\begin{figure}[tbp]
\begin{center}
\epsfig{file=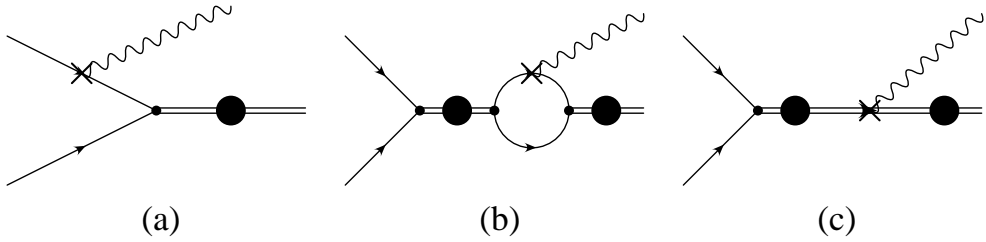,width=13cm}
\caption{\label{fig;npdg}
Diagrams contributing to $np\to d\gamma$:
Diagrams (a) and (b) are of $\calO(Q^{1/2})$,
while  diagram (c) is 
of $\calO(Q^{3/2})$.
The single lines,
the double lines with a filed circle, 
and the wavy lines
denote nucleons, dibaryons, and photons, respectively.
The double line with a filed circle,
which is obtained in Fig.~\ref{fig;prop-d} in 
Appendix,
represents the propagation of $^3S_1$ and $^1S_0$ states
in the intermediate state 
and the wave function normalization factor
$\sqrt{Z_d}$ in the final state.
}
\end{center}
\end{figure}

Diagrams for the $np\to d\gamma$ reaction
are depicted in Fig.~\ref{fig;npdg}.
The diagrams (a) and (b) 
in Fig.~\ref{fig;npdg} 
are of LO, ${\cal O}(Q^{1/2})$,
and the diagram (c) in Fig.~\ref{fig;npdg} 
is of $\calO(Q^{3/2})$, thus of a higher order.
However, we will discuss below that,
because of the $Vdd$ vertex involved,
it would be appropriate
to extract LO contribution from the (c) diagram, which is of
the same order as that of the (a) and (b) diagrams.
Summing up the contributions 
of the diagrams (a), (b) and (c)
in Fig. \ref{fig;npdg},
one obtains the amplitudes
for the initial $^3S_1$ and $^1S_0$ states as
\bea
i{\cal A}^{(a+b+c)}({}^3S_1) &=&
-i(\vec{\epsilon}_{(d)}^*\times\vec{\epsilon}_i)
\cdot (\vec{\epsilon}_{(\gamma)}^*\times \hat{k})
\sqrt{\frac{2\pi\gamma}{1-\gamma\rho_d}}
\frac{2}{m_N}
\frac{1}{-\gamma-ip+\frac12 \rho_d(\gamma^2+p^2)}
\nnb \\ && \times \left[
\frac{1+\kappa_S}{2m_N}\frac12 \rho_d(\gamma^2+p^2)
+\frac{\gamma^2+p^2}{m_N}L_2\right] ,
\label{eq;A3S1}
\\
i{\cal A}^{(a+b+c)}({}^1S_0) &=&
\vec{\epsilon}_{(d)}^*\cdot
(\hat{k}\times\vec{\epsilon}_{(\gamma)}^*)
\sqrt{\frac{2\pi\gamma}{1-\gamma\rho_d}}
\frac{2}{m_N}
\frac{1}{-\frac{1}{a_0}-ip+\frac12 r_0p^2}
\nnb \\ && \times
\left[
\frac{1+\kappa_V}{2m_N}\left(
\gamma-\frac{1}{a_0}+\frac12 r_0 p^2
\right)
+ \frac{\gamma^2+p^2}{2m_N} L_1
\right].
\label{eq;A1S0}
\eea
Here $\vec{p}$ is the relative momentum of the two-nucleon
system, and $\vec{k}$ is the momentum of the out-going photon;
$p=|\vec{p}|$,
$k=|\vec{k}|$,
and $\hat{k}=\vec{k}/k$.
$\vec{\epsilon}_{(d)}^*$ and $\vec{\epsilon}_{(\gamma)}^*$
are the polarization vectors for the out-going deuteron
and photon, respectively, and
$\vec{\epsilon}_i$ is the spin $S=1$ vector for
the initial ${}^3S_1$ state.
$a_0$ and $r_0$ are, respectively,  
the scattering length and effective range of
the neutron--proton $^1 S_0$ state.
The LEC $L_2$ has already been fixed
in the previous section
using the deuteron magnetic moment $\mu_M$.
The LEC $L_1$ should be fixed 
by experiment.

It is worth noting that
the LO (${\cal O}(Q^{1/2})$) amplitude
for the ${}^3S_1$ state in Eq.~(\ref{eq;A3S1}) vanishes
with the modified counting of $L_2$,
\bea
\left. {\cal A}^{(a+b+c)}(^3S_1) \right|_{L_2=L_2^0} = 0 \, ,
\eea
where the expression of $L_2^0$ is given in Eq.~(\ref{eq;L20}).
It is well known that the 
leading isoscalar transition
between the scattering and bound states
vanishes due to the orthogonality of the
wave functions. 
We confirm that
our treatment of the LEC $L_2^0$
satisfies this condition.
Higher-order terms can give non-vanishing
contribution to the isoscalar amplitude,
but since they are very small compared to the 
dominant isovector one \cite{pkmr-plb00},
we neglect them 
in following calculations.

We determine the value of $L_1$
in Eq.~(\ref{eq;A1S0})
from the cross section of $np\to d\gamma$ at the thermal energy,
$\sigma^{exp}=334.2\pm 0.5$ mb \cite{cwc-npa65}.
The total cross section in the CM frame reads
\bea
\sigma =
\frac{\alpha(\gamma^2+p^2)}{4 p}
\sum_{spin}|{\cal A}|^2 ,
\eea
where $\alpha$ is the fine structure constant,
and the CM energy $E$ corresponding to the thermal neutron
experiment
is $E=p^2/m_N=1.264\times 10^{-8}$ MeV.
Using the formula for the spin summation
$\sum_{spin}|i\vec{\epsilon}_{(d)}^*\cdot
(\hat{k}\times\vec{\epsilon}_{(\gamma)}^*)|^2 = 2$
and including only the amplitude for the $^1S_0$ channel,
${\cal A}(^1S_0)$ in Eq.~(\ref{eq;A1S0}),
we obtain $L_1 = -4.427\pm 0.015$ fm.

It is to be noted that,
if $L_1$ is set equal to zero, {\it i.e.}, $L_1=0$,
the relevant cross section would become
$\sigma(L_1=0) = 502.3$ mb,
which is about 1.5 times larger than
the experimental value, {\it i.e.},
$\sigma(L_1=0)/\sigma^{exp} \simeq $ 1.50.
The magnitude of the LO cross section
in the KSW scheme (Eq.~(36) in Ref.~\cite{ssw-npa99})
is smaller than $\sigma^{exp}$
by about 13 \%. 
Furthermore,
the leading one-body operators in the potential
model calculations also lead to a cross section
that is smaller than $\sigma^{exp}$
by about 10 \%.
As is well known, most of this 10 \% deficiency 
can be accounted for by the meson exchange currents.
Thus, here again we are facing the situation
that, whereas the conventional treatments
indicate the dominance of the LO contributions,
the dEFT results exhibit uncomfortably large
higher order corrections.
To solve this problem, we assume
that $L_1$ is dominated by a leading contribution
denoted by $L_1^0$ which is chosen so as to
reproduce the result of ERT, and that the rest, $\delta L_1$
contains information about the high energy physics
that has been integrated out.
Thus we consider the decomposition
\bea
L_1 = L_1^0 +\delta L_1, \ \ \
L_1^0= -\frac14 (1+\kappa_V)(r_0+\rho_d) ,
\label{eq;L10}
\eea
where numerically $L_1^{0} = -5.275$ fm.
If we insert the expression for $L_1^0$ into $L_1$ in the lagrangian,
the coefficient of the $Vdd$ term becomes
$-\frac{1+\kappa_V}{2m_N}\frac{r_0+\rho_d}{2\sqrt{r_0\rho_d}}$,
which is the coefficient of the one-body $VNN$ interaction
apart from a factor $(r_0+\rho_d)/(2\sqrt{r_0\rho_d})\simeq 1.024$
(which is very close to unity).
Since $L^0_1$ is ${\cal O}(Q^{-1})$
due to the factor $(r_0+\rho_d)$
in Eq.~(\ref{eq;L10}),
the order of diagram (c) in Fig.~\ref{fig;npdg} 
becomes the same as that of diagrams (a) and (b).
At LO, the total cross section reads
\bea
\sigma(L_1^0) &=&
\frac{\pi\alpha(1+\kappa_V)^2\gamma^5a_0^2}{pm_N^4(1-\gamma\rho_d)}
\left[
1
-\frac{1}{\gamma a_0}
-\frac{\gamma}{4}(r_0+\rho_d)
\right]^2 \, ,
\label{eq;sigmaER}
\eea
which is the same expression as that of ERT.
Numerically we have
$\sigma(L_1^0)=304.9$ mb.
This value is very close to
$\sigma_{IA}=304.5$ mb
obtained in the potential model
with the use of the one-body operator.

\begin{figure}[tbp]
\begin{center}
\epsfig{file=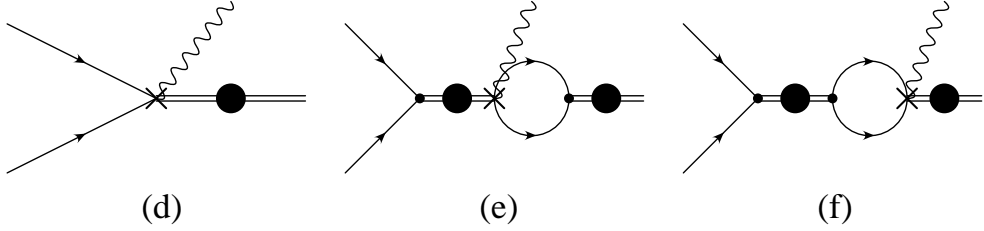,width=13cm}
\caption{\label{fig;npdg-L1p}
Higher order diagrams for the reaction $np\to d\gamma$
involving the $VdNN$ vertex
proportional to $L_1'$.
The order of diagrams is of $\calO(Q^{5/2})$.
See the caption of Fig.~\ref{fig;npdg} as well.
}
\end{center}
\end{figure}
Now we study higher order corrections
for the $np\to d\gamma$ reaction.
Higher order diagrams (${\cal O}(Q^{5/2})$) 
with the $VdNN$ vertex proportional to $L_1'$
are depicted in Fig.~\ref{fig;npdg-L1p}.
One may notice 
that the diagrams with the $D$-wave contribution vanish
for the initial $S$-wave states and
the corrections from
the radii are negligible
because of tiny momentum transfer in the reaction.
The combined contributions of diagrams (d,e,f) in Fig.~\ref{fig;npdg-L1p}
are given by
\bea
i{\cal A}^{(d+e+f)}(^1S_0) &=&
\vec{\epsilon}_{(d)}^*\cdot \left(
\hat{k}\times \vec{\epsilon}_{(\gamma)}^*\right)
\sqrt{\frac{2\pi \gamma}{1-\gamma\rho_d}}
\frac{2}{m_N}
\frac{1}{-\frac{1}{a_0}-ip+\frac12r_0p^2}
\nnb \\ && \times
\frac{\gamma^2+p^2}{2m_N}
\sqrt{\frac{m_N}{2\pi}}
\left(\gamma+\frac{1}{a_0}
-\frac12r_0p^2\right) L_1' \, .
\label{eq;A1S0def}
\eea
Comparing 
the amplitude
in Eq.~(\ref{eq;A1S0def}) with that in Eq.~(\ref{eq;A1S0}),
one finds that the momentum dependences
of the amplitudes for the LEC's $L_1$
and $L_1'$
are slightly different.
This difference, however, does not affect
the numerical results at small energies
significantly.
Here we compute the M1 transition contribution to
the $np\to d\gamma$ cross section,
using two sets of the parameters:
$\delta L_1=0.736$ fm and $L_1'=0$,
and $\delta L_1 = 0$ and $L_1'=4.464$ fm$^{5/2}$;
these LEC's have been fixed from $\sigma^{exp}$.
Predictions for the 
$np\to d\gamma$ cross section
at energies up to 1 MeV above
the thermal neutron energy
are given in Table~\ref{table;sigmaM1}.
A fairly good agreement is seen between
the results corresponding to the two different choices 
of the parameter set.
Our dEFT results also show reasonable agreement
with the recent EFT calculations~\cite{cs-prc99,r-npa00},
as well as with the potential model calculation
including the exchange current~\cite{nsgk}.
\begin{table}[tbp]
\begin{center}
\begin{tabular}{ c | c  c  c  c }\hline
E(MeV)              & $\sigma_{M1}(\delta L_1)$ & 
$\sigma_{M1}(L_1')$ & $\sigma_{M1}(EFT)\cite{r-npa00}$ &
$\sigma_{M1}(Av18)$\cite{nsgk}\\ \hline
$1.264 \times 10^{-8}$ & 334.2$^*$ & 334.2$^*$ & 334.2$^*$ & 334.2$^*$ \\
$5\times 10^{-4}$      & 1.668 & 1.668 & 1.667 & 1.668 \\
$1\times 10^{-3}$      & 1.171 & 1.171 & 1.170 & 1.171 \\
$5\times 10^{-3}$      & 0.4953 & 0.4954 & 0.4950 & 0.4954 \\
$1\times 10^{-2}$      & 0.3281 & 0.3281 & 0.3279 & 0.3281 \\
$5\times 10^{-2}$      & 0.09814 & 0.09820 & 0.09810 & 0.09820 \\
$0.100$                & 0.04969 & 0.04976 & 0.04973 & 0.04975 \\
$0.500$                & 0.00775 & 0.00782 & 0.00787 & 0.00781 \\
$1.00$                 & 0.0035  & 0.0036  & 0.0036  & 0.00355  \\\hline 
\end{tabular}
\caption{
M1 transition contribution to the cross section 
for the $np\to d\gamma$ reaction.
The second and third columns are our dEFT results
using the two sets of LEC's described in the text.
The fourth column shows the results
of the NNLO EFT calculation~\cite{r-npa00},
while the fifth column gives the results of
an accurate potential model calculation (Av18)
including the exchange current~\cite{nsgk}.
The LEC's in both dEFT and EFT
and the strength of the exchange current in the potential
model calculation are fixed
at the values marked by *.
}\label{table;sigmaM1}
\end{center}
\end{table}

\vskip 3mm \noindent
{\bf 5. Discussion and conclusions}

In this paper
we have reexamined dEFT
without pions for the deuteron reactions involving
the EM probe.
The EM form factors of the deuteron and
the total cross sections of
the $np\to d\gamma$ reaction have been calculated.
We have found that the LEC's of the $Vdd$ vertices in dEFT
give much larger contributions than
the corresponding terms in 
the previous EFT approaches.
Although it has not been completely
explored
why the LEC terms of the $Vdd$ vertex
give such large contributions in dEFT,
we have proposed a practical prescription to 
extract the LO part of the LEC's
in such manner that this LO part
mimics the leading one-body $VNN$ vertices.
After the LO part is 
disentangled, the remainder 
is assumed to represent
the high energy physics that has been integrated out.
If one is to calculate it explicitly, one
should take into account
relativistic corrections, meson-exchange currents,
higher partial waves, {\it etc}.
However, an easy way to fix
the higher order part of the LEC's 
is to fit it to experimental data. 
We were able to confirm that
the higher order corrections defined in this way
are indeed small,
in conformity with the general tenet of EFT.

With the LEC's thus determined, 
we have calculated the form factors $A(q)$ and $B(q)$
for elastic $e$-$d$ scattering
and found that the results 
of $A(q)$ 
are very close to
the experimental data for momenta up to 200 MeV.
Furthermore,
our results of $A(q)$ and $B(q)$
are comparable to
those of the accurate potential model calculation.
Our estimations of the total cross sections
of the $np\to d\gamma$ reaction for
energies up to 1 MeV also agree well with
the results of the other EFT calculations
and the accurate potential model.
With the proper treatment of the LEC's in the $Vdd$ vertices,
the convergence of dEFT becomes 
similarly efficient as other EFT's
and this can be interpreted as an indication
that dEFT is a useful tool for understanding 
a wide class of low-energy phenomena involving the deuteron
and an external probe.

\vskip 3mm \noindent
{\bf Acknowledgments}

We thank S. Nakamura
for providing us his numerical results
in Table 1 and communications.
We also thank K. Kubodera
for reading the manuscript.
S.A. thanks
U. van Kolck,
M. Oka,
R.~H. Cyburt,
W. Detmold,
J.-W. Chen,
A. Parreno,
P.~F. Bedaque,
S.~R. Beane,
T. Sato, 
F. Myhrer,
H.~W. Fearing,
and M.~J. Savage
for discussions and communications.
This work is supported in part
by the Natural Sciences and Engineering
Research Council of Canada
and by the Korea Research Foundation
(Grant No. KRF-2003-070-C00015).



\vskip 3mm \noindent
{\bf Appendix}

In this appendix we rederive
the two-nucleon(dibaryon) propagators and 
fix the LEC's using 
the known properties of
$NN$ systems in the $S$-wave channels;
see also Ref.~\cite{bs-npa01}.
 
\begin{figure}[tbp]
\begin{center}
\epsfig{file=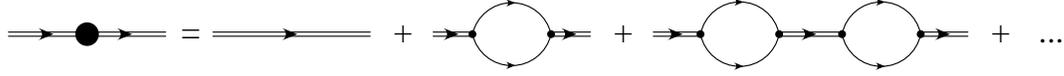,width=14cm}
\caption{\label{fig;prop-d}
Diagrams for the two-nucleon(dibaryon) propagator at leading order:
a double (single) line stands for the dibaryon (nucleon) field.}
\end{center}
\end{figure}

LO diagrams for the two-nucleon(dibaryon) propagators
in the $S$-wave channels
are depicted in Fig.~\ref{fig;prop-d}.
Since the insertion of the two-nucleon one-loop diagram
does not alter the order of the 
diagram~\cite{ksw-plb98,vk-npa99,g-98,bk-plb98},
the two-nucleon bubbles in the propagators
should be summed up to infinite order.
Thus the inverse
two-nucleon(dibaryon) propagators
in the center-of-mass (CM) frame read
\bea
i D_{t,s}^{-1}(p) &=& 
i\sigma_{t,s}(E+\Delta_{t,s})+iy_{t,s}^2\frac{m_N}{4\pi}(\mu+ip)
\nnb \\ &=&
i\frac{m_Ny_{t,s}^2}{4\pi}\left[
\frac{4\pi\sigma_{t,s}\Delta_{t,s}}{m_Ny_{t,s}^2}
+\mu
+\frac{4\pi\sigma_{t,s} E}{m_Ny_{t,s}^2}
+ip\right],
\label{eq;d-prop}
\eea
where $\mu$ is the renormalization scale 
of the PDS scheme, 
$p$ is the magnitude of the nucleon momentum
in the CM frame,
and $E$ is the total energy $E\simeq p^2/m_N$.

The two-nucleon(dibaryon) propagator in the spin singlet state,
$D_s(p)$, can be renormalized using the 
${}^1S_0$ channel
$NN$ scattering amplitude.\footnote{
Although it is known that
the expansion series of the ERT parameters in the ${}^1S_0$ channel
converges well, we employ 
the modified counting rule
$Q\sim \{p,a_0^{-1},r_0^{-1}\}$ for the ${}^1S_0$ channel as well.
}
\begin{figure}[tbp]
\begin{center}
\epsfig{file=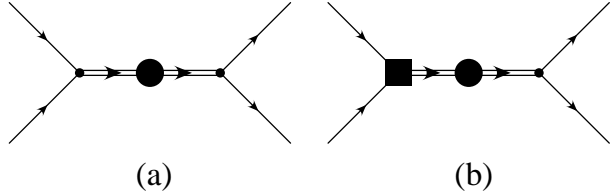,width=8cm}
\caption{\label{fig;NNamps}
Diagrams for the $NN$ scattering amplitudes.
Diagram (a) is for $S$- to $S$-wave channel
and diagram (b) is for $D$- to $S$-wave channel.
The $dNN$ vertex with (without) 
a filled box denotes a vertex 
proportional to $C_2^{(sd)}$ ($y_{t,s}$).
The propagator of the two-nucleon(dibaryon) filed 
(a double line with a filled circle) is obtained
from the diagrams in Fig.~\ref{fig;prop-d}.
}
\end{center}
\end{figure}
The amplitude obtained from Fig.~\ref{fig;NNamps} (a) reads
\bea
i{\cal A}_{s} &=& (-iy_{s})(iD_{s}(p))(-iy_{s})
= \frac{4\pi}{m_N}\frac{1}{
-\frac{4\pi\sigma_s\Delta_{s}}{m_Ny_{s}^2}
-\mu
-\frac{4\pi\sigma_s}{m_N^2y_{s}^2}p^2
-ip},
\eea
and it is related to the $S$-matrix via
\bea
S-1 = e^{2i\delta_0}-1
= \frac{2ip}{p\,{\rm cot}\delta_0-ip}
= i\left(\frac{pm_N}{2\pi}\right){\cal A}_s,
\label{eq;S}
\eea
where $\delta_0$ is the phase shift
for the $^1S_0$ channel.
Meanwhile effective range expansion (ERE) reads
\bea
p\,{\rm cot}\delta_0 &=&
-\frac{1}{a_0} +\frac12 r_0 p^2 + \cdots,
\label{eq;ERT}
\eea
where $a_0$ (= $-$23.71 fm) is the scattering length and
$r_0$ (= 2.73 fm) the effective range in the $^1S_0$ channel.
Inserting Eq.~(\ref{eq;ERT}) into Eq.~(\ref{eq;S}),
one obtains $\sigma_s = -1$, and 
\bea
y_s = \frac{2}{m_N}\sqrt{\frac{2\pi}{r_0}},
\ \ \ 
D_s(p) = 
\frac{m_N r_0}{2}
\frac{1}{\frac{1}{a_0}+ip -\frac12r_0p^2}\, .
\label{eq;1s0ren}
\eea

Now we fix the LEC's in the coupled channel.
Diagrams for the $^3S_1-{}^3 D_1$ channel
are depicted in Figs.~\ref{fig;NNamps} (a, b).
Each of the amplitudes reads
\bea
{\cal A}_{[SS]} &=&
\frac{4\pi}{m_N}\frac{1}{
-\frac{4\pi\sigma_t\Delta_{t}}{m_Ny_{t}^2}
-\mu
-\frac{4\pi\sigma_t}{m_N^2y_{t}^2}p^2
-ip}\, ,
\\
{\cal A}_{[DS]} &=&
\frac{\sqrt{2}}{3}\frac{C_2^{(sd)}p^2}{y_t\sqrt{m_N\rho_d}} 
\frac{4\pi}{m_N}
\frac{1}{
-\frac{4\pi\sigma_t\Delta_t}{m_Ny_t^2}
-\mu
-\frac{4\pi\sigma_t}{m_N^2y_t^2}p^2
-ip}\, .
\eea
The relation between the $S$-matrix and
the amplitudes in the coupled channel 
is given by
\bea
S&=& 1+i\frac{pm_N}{2\pi}\left(
\matrix{
{\cal A}_{[SS]} & {\cal A}_{[DS]} \cr
{\cal A}_{[SD]} & {\cal A}_{[DD]}}
\right)
= \left(\matrix{
 e^{2i\bar{\delta}_0}{\rm cos}2\bar{\epsilon}_1 & 
 ie^{i(\bar{\delta}_0+\bar{\delta}_2)}{\rm sin}2\bar{\epsilon}_1 \cr
 ie^{i(\bar{\delta}_0+\bar{\delta}_2)}{\rm sin}2\bar{\epsilon}_1 &
 e^{2i\bar{\delta}_2}{\rm cos}2\bar{\epsilon}_1
}\right),
\eea
where we have employed a convention for the phase shifts
defined in Ref.~\cite{sym-pr57}.

Since it is known that $\bar{\epsilon}_1$ 
is numerically small,
we put $\cos (2 \bar{\epsilon_1}) \simeq 1$.
Around the deuteron pole, ERE reads~\cite{b-pr49},
\bea
p\,{\rm cot}\bar{\delta}_0 = 
-\gamma
+\frac12\rho_d(p^2+\gamma^2)
+\cdots,
\eea
with
$\gamma^{-1} = 4.319$ fm and
$\rho_d=1.764$ fm.
Following the same steps that lead to Eq.~(\ref{eq;1s0ren}),
one obtains $\sigma_t=-1$, and
\bea
&& y_t = \frac{2}{m_N}\sqrt{\frac{2\pi}{\rho_d}},
\ \ \ 
D_t(p) =
\frac{m_N\rho_d}{2}\frac{1}{\gamma+ip-\rho_d\frac12(\gamma^2+p^2)}
= \frac{Z_d}{E+B} + \cdots,
\label{eq;Dt}
\eea
where $Z_d$ is the wave function normalization factor
of the deuteron around the pole of the deuteron binding energy $B$,
and the ellipsis in Eq. (\ref{eq;Dt})
denotes corrections that are finite or vanish at $E=-B$.
Thus one obtains~\cite{bs-npa01}
\bea
Z_d = \frac{\rho_d\gamma}{1-\rho_d\gamma}\, .
\label{eq;Zd}
\eea

Next we renormalize the coupled channel amplitude ${\cal A}_{[DS]}$
at the deuteron pole $p=i\gamma$
through the relation,
\bea
-2\left(\frac{\eta_{sd}}{1-\eta_{sd}^2}\right) =
\left.\frac{{\rm tan}(2\bar{\epsilon}_1)}
{{\rm sin}(\bar{\delta}_0-\bar{\delta}_2)}\right|_{p=i\gamma}
=\left.\frac{2{\cal A}_{[SD]}}{{\cal A}_{[SS]}-{\cal A}_{[DD]}}
\right|_{p=i\gamma}
\simeq -\frac13\sqrt{\frac{m_N}{\pi}}C_2^{(sd)}\gamma^2,
\eea
where we have neglected the $D$-wave amplitude
${\cal A}_{[DD]}$.
Thus, neglecting the $\eta_{sd}^2$ correction 
in the relation above, one obtains
\bea
C_2^{(sd)} &\simeq& 6 \sqrt{\frac{\pi}{m_N}}
\frac{\eta_{sd}}{\gamma^2}\, .
\label{eq;C2sd}
\eea

\end{document}